\documentstyle[prl,aps,multicol,epsfig]{revtex}

\begin{document}
\draft
\title{Vibrational origin of the fast relaxation processes in molecular glass-formers}

\author{
        S.~Mossa$^{1,2}$,
        G.~Monaco$^{3}$,
        and G.~Ruocco$^{2}$
                }
\address{
         $^1$
         Center for Polymer Studies and Department of Physics,
         Boston University, Boston, Massachusetts 02215  \\
         $^2$
         Dipartimento di Fisica and INFM, Universit\`a di Roma "La Sapienza"
         P.zza Aldo Moro 2, Roma, I-00185, Italy \\
         $^3$
         European Synchrotron Radiation Facility,
         BP220, Grenoble Cedex, F-38043, France \\
        }

\date{\today}
\maketitle
\begin{abstract}
We study the interaction of the relaxation processes with the density
fluctuations by molecular dynamics simulation of a flexible molecule model for
o-terphenyl (oTP) in the liquid and supercooled phases. 
We find evidence, besides the structural
relaxation, of a secondary {\it vibrational} relaxation whose
characteristic time, few ps, is slightly temperature dependent.
This i) confirms the result by Monaco et al. [Phys. Rev, E{\bf
62}, 7595 (2000)] of the vibrational nature of the fast relaxation
observed in Brillouin Light Scattering (BLS) experiments in oTP;
and ii) poses a caveat on the interpretation of the BLS spectra
of molecular systems in terms of a purely center of mass dynamics.
\end{abstract}


\begin{multicols}{2}

After the development of the Mode Coupling Theory
(MCT)~\cite{mct1}, aimed to give a microscopic interpretation to
the slowing down of the dynamics which takes place at the liquid-to-glass
transformation, different experimental and numerical studies have
been devoted to test its predictions, especially
through the study of the density fluctuations correlation
function, or of its space Fourier transform $F(Q,t)$. 
These studies can be assigned to two broad
classes, according to the value of the investigated wavevector $Q$. 
On one side the numerical techniques and the
inelastic neutron scattering spectroscopies, that access the
high $Q$ region, give a picture where the MCT
predictions are basically demonstrated. 
On the other side the low $Q$ techniques as, for
example, Brillouin Light Scattering (BLS), depolarized light
scattering, or dielectric spectroscopy, depict a much less clear
situation. For example, the recent observation of a flat
background (constant loss) in the dielectric absorption spectra
at few GHz~\cite{dielectric}, seems to contradict the MCT
predictions on the same systems where the high $Q$ techniques
confirm these predictions. A similar conflictual
situation is also found in the case of the BLS
experiments~\cite{light_scatt}, in particular as far as the so
called $\beta$-region of the MCT is concerned. This is the time
region where the first decay of $F(Q,t)$, that could be ascribed to the
dephasing of the microscopic vibrational dynamics, merge with the
earliest part of the decay associated to the structural
($\alpha$) relaxation process. The MCT makes specific predictions
on the shape of the $F(Q,t)$ in this region, but these predictions
are not in agreement with the BLS, low $Q$, measurements
\cite{BLSnoMCT} whereas the high $Q$ neutron spectroscopy
\cite{neutroni} and molecular dynamics \cite{MD} data support the
validity of the MCT.

Recently, studying the BLS spectra~\cite{monfiomasc} 
of o-terphenyl (oTP), one of
the prototypical fragile glass formers, 
some of us found evidences -besides the usual structural or
$\alpha$ relaxation process- of a secondary relaxation, with a
characteristic time $\tau_f$ lying in the ten picosecond
time-scale and weakly temperature dependent. This process 
lies in the frequency
region where the features associated to the $\beta$ region are
expected to show up. By comparing the BLS spectra in the glassy
and in the ordered crystalline phases, and by comparing the
effect of this relaxation process on the longitudinal and
transverse sound waves, it has been suggested the {\it
vibrational} nature of the observed relaxation~\cite{moncapdile}.
If confirmed, this suggestion would indicate the reason why the
MCT predictions are not always retrieved in the light scattering
and dielectric experiments. Indeed the MCT is, up to now,
formulated for monatomic or rigid
molecules, the complexity of real non-rigid molecules -with all
the processes associated to the internal vibrational dynamics-
not being taken into account. The presence of a
vibrational relaxation would indicate that the molecular
glass formers are not suitable benchmarks to test the MCT
on the short time scale. Indeed
the presence of a vibrational relaxation dynamics lying in the
1-100 ps time-scale would mask the spectral features predicted in
the $\beta$-region. This considerations could explain
the lack of consistency between the MCT predictions and the low
$Q$ experiments.

In this letter we present a molecular dynamics simulation
investigation of the relaxation processes active in oTP in a wide
temperature range, spanning the normal and supercooled liquid
region. Using the recent flexible molecular model for oTP~\cite{otp1,otp2}, 
we investigate the coupling between the density
fluctuations and the intra-molecular (vibrational) degrees of
freedom. We find that i) there is a strong coupling between the
density fluctuations and the vibrational excitations, and ii)
that this coupling gives rise to a vibrational relaxation in the
10$^{-11}$ s time-scale. 

These findings strongly support the
hypothesis suggested in Ref.~\cite{moncapdile} and confirm the
existence of a vibrational relaxation in oTP in the same
frequency region where the signatures of the MCT $\beta$-region
are expected.
\begin{figure}
\centering
\epsfig{file=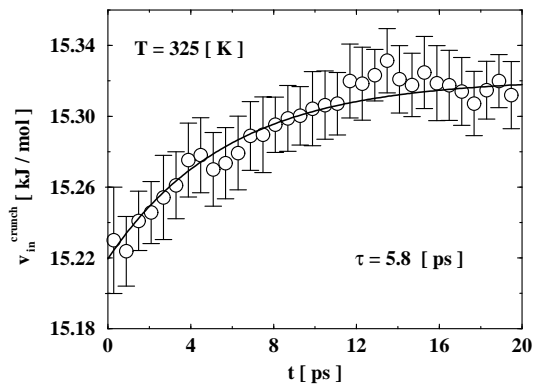,width=.8\linewidth,angle=0}
\caption{Time dependence of the total intra-molecular potential
energy after the density jump starting from a well equilibrated
configuration at $T=325$ K. The data have been obtained averaging
over 2,000 realizations and the error bars represent the
$\pm\sigma$ standard deviation of the different runs data.}
\label{fig1}
\end{figure}
\noindent
Moreover, by analyzing the effects of an external
perturbation on the individual contributions to the
intra-molecular interaction potential, we identify the
phenyl-phenyl stretching as the internal vibration mainly
responsible for such a relaxation. 

The absolute value of the
relaxation time $\tau_f$ turns out to be a factor $\approx$ 3
shorter than the experimental value, a discrepancy that is
tentatively explained on the basis of the
classical nature of the MD simulations. Finally, we investigate
the temperature dependence of $\tau_f$, finding a behavior very
similar to that observed in the BLS experiment~\cite{otp_pap}.

We have studied a microcanonical system composed
by 32 molecules (96 rings, 576 Lennard Jones interaction sites)
enclosed in a cubic box with periodic boundary conditions.
In the utilized  model~\cite{otp1} the oTP molecule is constituted
by three rigid hexagons (phenyl rings) of side $L_{a}=0.139$ nm.
Two adjacent vertices of the parent ring are bonded to one vertex
of the two lateral rings by bonds of equilibrium length
$L_{b}=0.15$ nm. Each vertex of the hexagons is occupied by a
fictious atom of mass $M_{CH}=13$ a.m.u. representing a
carbon-hydrogen pair (C-H). In the isolated molecule equilibrium
position the lateral rings lie in planes that form an angle of
about 54$^o$ with respect to the parent ring's plane. The
interaction among the rings pertaining to all the molecules
is described by a site-site pairwise additive
Lennard-Jones 6-12 potential, each site corresponding to the six
hexagons vertices. Moreover, the three rings of a given molecule interact
among themselves by an {\it intra-molecular} potential, such
potential being chosen in such a way to preserve the molecule
from "dissociation", to give the correct relative equilibrium
positions for the three phenyl rings, and  to represent at best
the isolated molecule vibrational spectrum. In particular it is
written in the form $V_{\mbox intra}=\sum_k c_k V_k$ where each
term $V_k$ controls a particular degree of freedom.
The contributions to the intra-molecular potential relevant to
the present study are: i) the stretching along
the central ring - side ring bonds ($S$), ii) the bending of the
central ring - side rings bonds ($B$), and iii) the in-phase ($R_1$)
and out-of-phase ($R_2$) rotation of the lateral rings along the
ring-ring bond~\cite{nota_pot}. 
The first two terms are modeled by springs, and
the "interaction" with the other degrees of freedom
(anharmonicity) is introduced by the site-site LJ potential of
different molecules. The third term, on the contrary, has been modeled
in more realistic way. In this case the relevant variables are the
two angles $\{\Phi_1,\Phi_2\}$ between the normals to the lateral
rings and the parent ring plane. An ab-initio calculation of the
single molecule potential energy surface as a function of these
two angles has shown that exist two iso-energetic configurations
separated by an energy barrier of height $V_s/k_B=580\,K$. The nature of the rotational motion
at the temperatures of interest can be summarized as follows: the
two side rings can pivot in phase around the bonds crossing from
one minimum to the other degenerate one; at the same time they
can perform librational out-of-phase motions. In order to
represent this potential surface we express the in-phase rotation
of the two side rings with a high-order (6th) polynomial and the
out-of-phase rotation by a quadratic (harmonic) potential energy.
Other details of the intra-molecular and inter-molecular
interaction potentials, together with the values of the involved
constants, are reported in Ref.~\cite{otp1}. Previous studies on
the temperature dependence of the self diffusion
coefficient~\cite{otp1} and of the structural ($\alpha$)
relaxation times~\cite{otp1,otp2} indicate that this molecular
model is capable to quantitative reproduce the dynamical
behavior of the real system, but the actual simulated
temperatures must be shifted by $\approx$ 20 K upward. In what
follows, as our aim is to compare the simulation results with the
experiments, the reported MD temperatures are always shifted by
such an amount. 

To investigate the effect of a long wavelength density
fluctuation, as those probed in a BLS experiment, on the
intra-molecular vibrational dynamics we proceed as follows. After
an equilibration run at a given temperature, we make a sudden
density variation of the system (``crunch''), then we follow the
time evolution of the intramolecular potential energy 
during the subsequent evolution.
As an example, in Fig.~\ref{fig1} we report such a time evolution
for $T$=325 K and averaged over 2,000
statistically independent crunches.
The value of the internal potential energy is close to that
pertaining to 12 harmonic oscillators (the number of internal
degrees of freedom), 6$RT$=16.2 kJ/mol, the slight deviation
being associated to the anharmonicity present in the
parameterizations of some of the internal degrees of freedoms. It
is clear that, after the density change, the vibrational energy
relaxes toward its new 
equilibrium value in a time-scale of
$\approx$6 ps, a value not far from the experimental one:
$\tau_f \approx$20 ps. We do not expect a much better
agreement between these two values as i) the intra-molecular
potential model has been parameterized to represent the vibrational
\begin{figure}
\centering
\epsfig{file=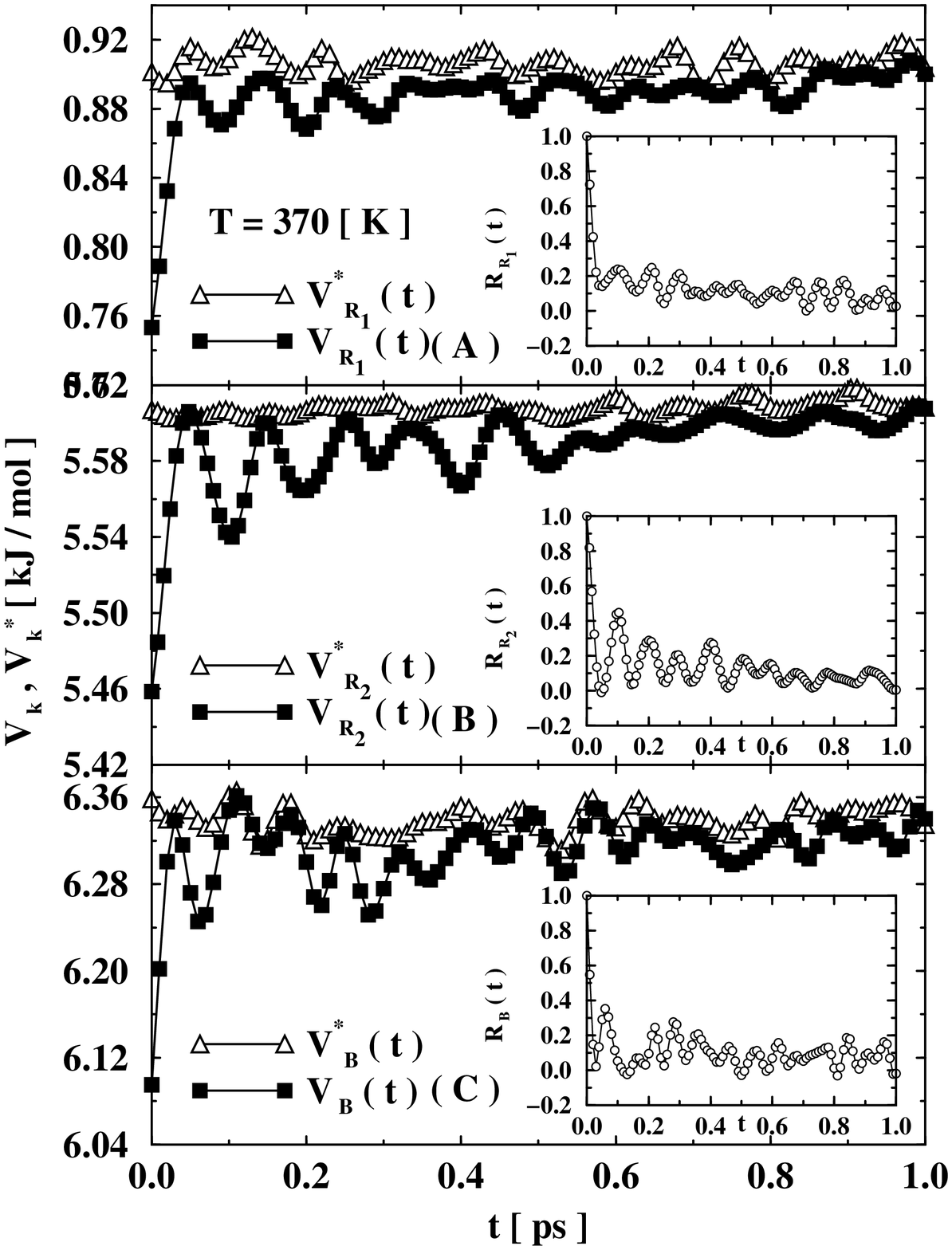,width=.8\linewidth,angle=0}
\epsfig{file=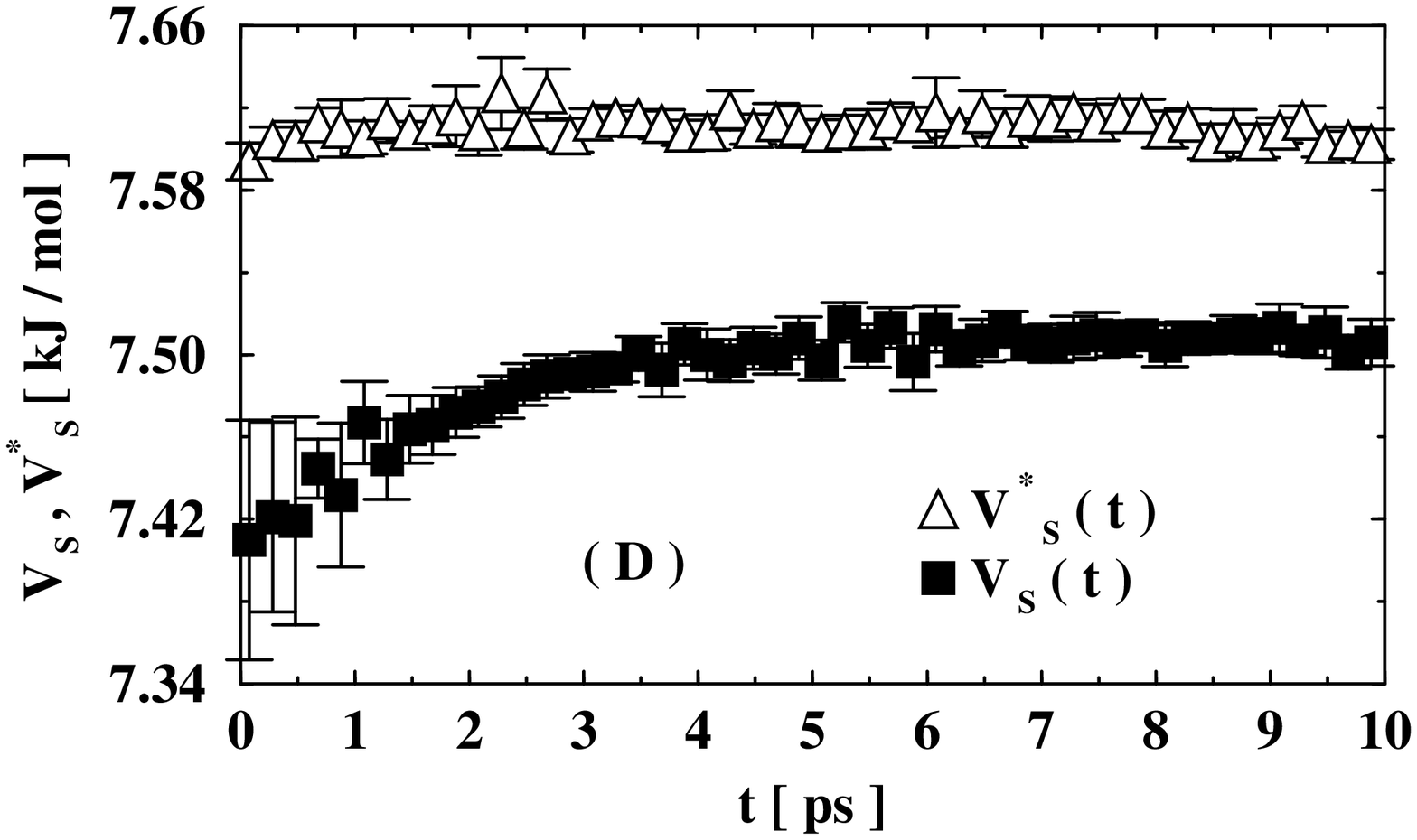,width=.8\linewidth,angle=0}
\caption{(A) Signals (main panel) and relaxation function (inset)
of the potential controlling the on-phase pivoting of the lateral
rings at $T=370$ K and $\lambda=1.2$. (B) As above for the
out-phase librations. (C) As above for the bending
of the central ring - side rings bond. 
(D) As above for the stretching between the sides
and central ring along the molecular bonds;
it is worth noting that $V_s$ decays on a time scale
much longer than that of the other terms shown above. 
} \label{fig2}
\end{figure}
\noindent
spectrum of the isolated molecules -and not on the
dynamics in the condensed phase-, and ii) the intramolecular vibrational
frequencies have values up to $\approx$500 cm$^{-1}$
($\approx$720 K) and therefore, at the investigated temperature,
the quantum effects (population of the vibrational levels) certainly
play a relevant role, and they are not considered in the present
classical simulation.

Overall, we conclude from Fig.\ref{fig1} that there is
definitively a coupling between the density fluctuations and the
vibrational degrees of freedom -i.~e. a vibrational relaxation is
active in the system- and that the relaxation time favorably compares 
with the experimental findings.

In order to identify which one of the 
internal degree of freedom(s) is more efficiently
coupled to the density fluctuations we cannot use the crunch
technique, as a 
\begin{figure}
\centering
\epsfig{file=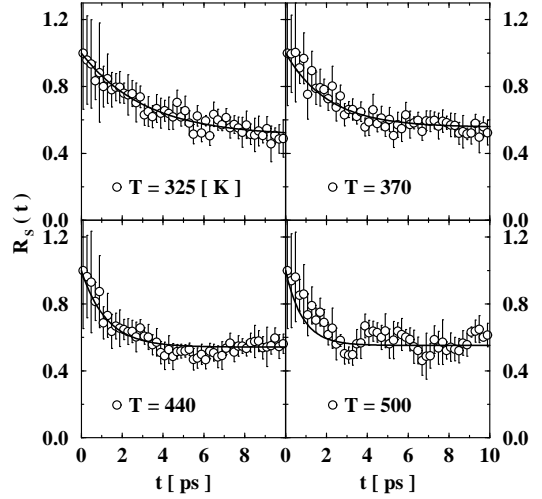,width=0.8\linewidth,angle=0}
\caption{Relaxation functions $R_S(t)$ of the bond stretching
potential at the four indicated temperatures.} \label{fig3}
\end{figure}
\noindent
density jump simultaneously affects all the
intra-molecular degrees of freedom.
In order to selectively
perturb a specific vibration, we proceed as follows. After the
equilibration run at a given temperature, the value of one of the
elastic constants $c_k$ is sudden scaled ($ c_k\rightarrow c_k^*
= \lambda c_k$)\cite{nota_c,nota_d} with $\lambda$ positive and
small~\cite{nota_linear}. After this perturbation is applied, we
follow the time evolution of the specific term of the
intra-molecular interaction potential energy and we measure the time
needed to the energy to relax towards its new equilibrium value.
As an example, in Fig~\ref{fig2} (A-D), we report the time
evolution of some of the terms $V_k$ ($k=R_1,R_2,B,S$) for the
selected temperature $T$=370 K (open squares). The value of the
energies measured in runs without the perturbation (open
triangles) are also shown for comparison. Finally, in the inset,
the difference $R_k(t)$ (normalized to unity at $t$=0) between the
perturbed and unperturbed time evolution are shown. As can be
seen, all the potential term but $V_S$ decay on a very fast
time-scale ($\approx$100 fs or less followed by an oscillating
tail that vanish in $\approx$0.5 ps). The term
$V_S$ on the contrary decays on a much longer ($>$ 1 ps) time scale.
This term, which controls the stretching along the two bonds
connecting the parent ring with the two side rings, is therefore
the main candidate for the vibrational relaxation observed at few
ten ps. It is worth noting that, at variance with the other
cases, the perturbed signal does not relax at the same value
pertaining to the equilibrium signal; this indicates the presence
of a relevant anharmonicity affecting this degree of freedom
and that its equilibrium value depends on the density.

\begin{figure}
\centering
\hspace{-0.8cm}
\epsfig{file=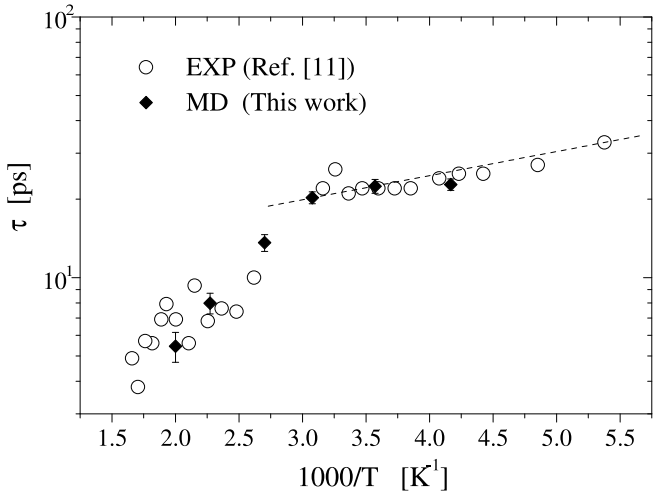,width=.75\linewidth,angle=-90.}
\vspace{3.mm} \caption{Experimental (open circles, From Ref. [11])
and MD (full diamonds) results for $\tau_f(T)$ reported in an
Arrhenius plot. The MD data have been multiplied by a common
factor $F=6$ in order to show the $T$ behaviour similar to that of the
experimental data. } \label{fig4}
\end{figure}

The functions $R_S(t)$ are reported in Fig.~\ref{fig3} at
selected temperatures, together with their best fit to an
exponential decay. A slight but evident temperature dependence 
of the relaxation time is present.
The $T$ dependence of the
relaxation time derived from these fits is reported in
Fig~\ref{fig4}, together with the relaxation times for the fast
\noindent process experimentally determined in Ref.~\cite{otp_pap}.
In this
figure the MD data have been multiplied  by a common factor $F=6$
in order to emphasize that their $T$ behaviour is very similar to
that of the experimental data. Both sets of data show an Arrhenius
behaviour at low $T$ (with activation energy $\Delta
E$=0.28$\pm$0.01 kJ/mole) and a different (steepest) trend at
high $T$, where the fast relaxation process merges with the
structural one. The agreement between the $T$ behaviour of the
two sets of data is remarkable. On the contrary, there is a
discrepancy of a factor of about 6 between the MD and the
experimental time-scales. This difference could be explained, as
noticed above, considering a non perfect parameterizations of the
intra-molecular interaction potential and the fact that we expect
strong quantum effects on the studied process. Moreover, in the
present case, further differences can arise from the different
"perturbation" utilized in the MD simulation (changes of force
constants) and in the experiment (density fluctuations).

In conclusion, by studying the coupling of external perturbations
(a density fluctuation and a fictious coupling constant
modification) with the intra-molecular vibrations in a flexible
model of oTP molecule we demonstrate i) the existence of a
vibrational relaxation process with a non negligible strength and
a relaxation time in the few ps time-scale; ii) that this
relaxation process is mainly associated to the phenyl-phenyl
stretching; and iii) that the relaxation time shows
a temperature behavior very similar 
to that of the experimentally determined
$\tau_f$~\cite{monfiomasc,moncapdile,otp_pap}. 
The quantitative discrepancies between the simulate and the
experimental relaxation times can be tentatively assigned to 
the quantum nature of the real vibration, for
which, at room temperature, $\hbar \omega_v/k_B T \approx 2$. The
present findings give strong support to the vibrational origin of
the fast relaxation process observed in oTP. It is tempting to
generalize this conclusion to other systems where such a fast
process has been observed (PC \cite{tesiRdL} and PB \cite{PB}).
Giving the depicted scenario, one should be extremely careful in
drawing conclusion on either the validity, or the failure, of the MCT
by the analysis of the light scattering or dielectric spectra. 
This is especially
true when analyzing the MCT $\beta$-region, as it lies at frequencies
that coincide with the typical fast process ones and as its main
features (susceptibility minima, $a$ and $b$ exponents, knees,
etc.) can be masked by the fast process itself.

We thank R.~Di~Leonardo and G.~Lacorata and F.~Sciortino for very useful
discussions.

\end{multicols}


\begin{references}
\bibitem{nota_pot}
An other contribution to the intramolecular potential is
taken into account in the model, i.e. the tilt
of the ring-ring bonds. Such terms are not
relevant in this context.
\bibitem{nota_c}
It is worth noting that the actual calculations have been
performed applying the modification $c_k^*\rightarrow c_k$
(instead of $c_k\rightarrow c_k^*$), after the equilibration of the
systems with $c_k^*$. In this way we always follow the relaxation
dynamics of the ``correct'' ($c_k$) system.
\bibitem{nota_d}
Actually, by modifying the elastic constants we are not
perturbing a specific vibration, as the latter are combinations of
the different internal variables (bonds lengths and angles).
However, as each molecular eigenmode mainly projects on one of the
internal variables, we can anyway identify the vibrational modes
responsible for the relaxation process.
\bibitem{nota_linear}
We have used a value of $\lambda=1.2$; this value has been chosen
by some preliminary run in such a way to observe clearly a
response to the perturbation introduced but to stay in the linear
response regime.
\end{references}
\end{document}